\documentclass{llncs}

\usepackage{amssymb}
\usepackage{amsmath}
\usepackage{graphicx}

\usepackage[OT4]{fontenc}

\usepackage[normalem]{ulem} 
\usepackage{soul} 

\usepackage{listings}

\newcommand{\mket}[1]{| #1 \rangle}

\newcommand{\mbraket}[2]{\langle #1 | #2 \rangle}

\newcommand{\imag}{\mathbf{i}}

\newcommand{\keywords}[1]{\par\addvspace\baselineskip\noindent\keywordname\enspace\ignorespaces#1}

\begin{document}


\mainmatter

\title{Quantum network protocol for qudits with use of quantum repeaters and Pauli Z-type operational errors}


\authorrunning{Marek Sawerwain \and Joanna Wi\'sniewska}
\tocauthor{Marek Sawerwain and Joanna Wi\'sniewska}

\author{Marek Sawerwain\inst{1} \and Joanna Wi\'sniewska\inst{2}}
\institute{Institute of Control \& Computation Engineering \\
University of Zielona G\'ora, Licealna 9, Zielona G\'ora 65-417, Poland \\
\email{M.Sawerwain@issi.uz.zgora.pl}
\and
Institute of Information Systems, Faculty of Cybernetics, \\
Military University of Technology, Kaliskiego 2, 00-908 Warsaw, Poland \\
\email{jwisniewska@wat.edu.pl}
}


\maketitle

\begin{abstract}
In this chapter a quantum communication protocol with use of repeaters is presented. The protocol is constructed for qudits i.e. the generalized quantum information units. One-dit teleportation is based on the generalized Pauli-Z (phase-flip) gate's  correction. This approach avoids using Pauli-X and Hadamard gates unlike in other known protocols based on quantum repeaters which were constructed for qubits and qudits. It is also important to mention that the repeaters based on teleportation protocol, described in this chapter, allow a measurement in the standard base (what simplifies the measurement process) and the use of teleportation causes only Pauli-Z operational errors.

\end{abstract}

\section{Introduction} \label{lbl:sec:introduction}

One of the solutions currently discussed in the field of quantum communication \cite{Stucki2011}, \cite{Ritter2012}, \cite{Nikolopoulos2014}  is a quantum repeater \cite{Briegel1998}. Its construction is not so simple as building classic amplifier because of non-cloning theorem which is one of the most basic and important foundations of quantum mechanics. However, the lack of ability of making perfect copies of quantum information may be compensated with use of teleportation and the phenomenon of entanglement. The protocol of quantum teleportation may be utilized to amplify the signal i.e. a quantum state during the transfer process. This approach may be used, e.g. in a fiber, to improve the quality of transferred information. Applying the teleportation protocol and the entanglement to amplify the quantum information results with building of so-called quantum repeater (QR).

It should be emphasized that the notion of QR is currently in the center of interest of many researchers  \cite{Meter2011}, \cite{Meter2014}. Apart from the theoretical analysis of this subject there are also physical experiments carried out with use of QRs as the elements of quantum networks. These experiments are being accomplished for the transmission of quantum states in an optical fiber \cite{Sangouard2011} and also in the air \cite{Ursin2007}. Therefore, the progress in the field of quantum computing \cite{Hirvensalo2001}, \cite{Klamka2004}, \cite{Klamka2001a}, \cite{Klamka2002}, \cite{Nielsen2001} is strongly connected with the progress of quantum communication because of the need to send information.

The notion of QR is currently discussed as a potential solution to the problems of quantum communication and cryptography (especially for quantum key distribution). The three generations of QR were presented do far. The first type utilizes the phenomenon of entanglement and its purification \cite{Briegel1998}, \cite{Duan2001}, \cite{Sangouard2011}. The second generation \cite{Jiang2009} is based on the near-perfectly entangled pairs' generation (however, in this approach still the entanglement purification is used). The third type \cite{Munro2012}, \cite{Muralidharan2014} of QR utilizes the protocol of teleportation. The quantum information is transmitted from one point to another, so the communication is organized as a one-way scheme. The solution presented in this chapter belongs to the third generation of QR and it uses so-called one-dit teleportation protocol to correct the errors which may appear during the transfer process. The novelty of described solution consists on using only Pauli-Z gate for error correction. More precisely: the result of measurement performed during the teleportation protocol determines the number of Pauli-Z operations which have to realized as an error correction.

The reminder of this chapter is organized as follows. In section (\ref{lbl:sec:single:dit:teleportation:MS:JW:CN:2016}) there is a quantum teleportation protocol presented and the error correction is performed with use of Pauli-Z gate. Section (\ref{lbl:sec:protocol:quantum:repeater:MS:JW:CN:2016}) describes the realization of QR with use of previously mentioned teleportation protocol. The interpretation of repeater protocol as a quantum circuit is shown in section (\ref{lbl:sec:qc:for:qr:network:MS:JW:CN:2016}). The summary and the final conclusions are presented in section (\ref{lbl:sec:conclusions:MS:JW:CN:2016}).

\section{Single dit teleportation protocol} \label{lbl:sec:single:dit:teleportation:MS:JW:CN:2016}

The protocols presented in this section are defined with notions of dits and qudits. A concept of dit is a generalization of classic bit. For classic bit there can be distinguished only two states: zero and one. In the case of dit more states are possible. The number of these values is symbolized by the letter $d$. For example, the classic bit is a dit with $d=2$.

A unit of quantum information is so-called qubit. A definition of qubit may be presented as:
\begin{equation}
\mket{\psi} = \alpha \mket{0} + \beta \mket{1}, \;\;\; {|\alpha|}^2 + {|\beta|}^2 = 1, \alpha, \beta \in \mathbb{C},
\end{equation}
where vectors $\mket{0}$, $\mket{1}$ stand for the computational base (in this case the standard base is used): 
\begin{equation}
\vert 0 \rangle =
\left[
\begin{array}{c}
1\\
0\\
\end{array}
\right],
\enspace
\vert 1 \rangle =
\left[
\begin{array}{c}
0\\
1\\
\end{array}
\right].
\end{equation}
These vectors represent the classic states zero and one.

By analogy, a generalization of qubit is a qudit. In this case more base states are admissible. The state of unknown qudit with $d$ base states is represented as:
\begin{equation}
\mket{\phi} = \alpha_0 \mket{0} + \alpha_1 \mket{1} + \alpha_2 \mket{2} + \ldots + \alpha_{d-1} \mket{d-1} , \;\;\; \sum_{i=0}^{d-1} {|\alpha_i|}^2 = 1, \alpha_i \in \mathbb{C}.
\end{equation}
The number of base states for a given qudit, expressed as $d$, will be also called a qudit's freedom level.

Standard base for qudits requires more vectors e.g. for so-called qutrits (qudits with $d=3$) standard basis vectors~are:
\begin{equation} 
\mket{0} = \left[ \begin{array}{c} 1 \\ 0 \\ 0\end{array} \right],\enspace \mket{1} = \left[ \begin{array}{c} 0 \\ 1 \\ 0 \end{array} \right],\enspace \mket{2} = \left[ \begin{array}{c} 0 \\ 0 \\ 1 \end{array} \right].
\end{equation}

The presented repeater protocol is realized as a single-dit teleportation protocol. The described protocol differs from other one-dit protocols \cite{Aliferis2004}, \cite{Knill2005} in this way that only Pauli-Z gate is used as a correction gate. The Pauli-Z (or just $Z$) gate for qudits may be defined as:

\begin{equation}
Z \mket{j} = \omega^{j}\mket{j},
\end{equation}
where $\omega$ stands for the root of unity:
\begin{equation}
\omega^{d}_{k} = \cos\left( \frac{2 k \pi }{d} \right) + \imag \sin\left( \frac{2 k \pi }{d} \right) = e^{\frac{2 \pi \imag k}{d}} , \;\;\; k=0,1,2,3, \ldots, d-1 .
\end{equation}
More precisely, $\omega^{d}_{k}$ is a $d$-th root number $k$ of unity. The symbol $\imag$ represents the imaginary unit.

There are also Hadamard gate and CNOT gate used. Definitions of these gates for qudits are: 
\begin{equation}
H \mket{j} = \sum_{k=0}^{d-1} \omega^{j \cdot k} \mket{k}, \;\;\; \mathrm{CNOT} \mket{ab} = \mket{a, a \overset{d}{\oplus} b}
\end{equation}
where $a \overset{d}{\oplus} b = (a + b) \mod d$. In the case of CNOT gate the $CNOT^{\dagger}$ gate may be given as well - this gate is obtained as the Hermitian conjugate of CNOT operation. The $CNOT^{\dagger}$ gate is useful for a quantum teleportation protocol described in this section.

The exemplary matrix form of CNOT gate for qutrits ($d=3$) is:
\begin{equation}
CNOT_3 = \left(\begin{array}{ccccccccc}
	1 & . & . & . & . & . & . & . & . \\
	. & 1 & . & . & . & . & . & . & . \\
	. & . & 1 & . & . & . & . & . & . \\
	. & . & . & . & 1 & . & . & . & . \\
	. & . & . & . & . & 1 & . & . & . \\
	. & . & . & 1 & . & . & . & . & . \\
	. & . & . & . & . & . & . & . & 1 \\
	. & . & . & . & . & . & 1 & . & . \\
	. & . & . & . & . & . & . & 1 & . \\
\end{array}\right)
\end{equation}
where the zeros were replaced by dots to make the notation more legible. In general, the CNOT gate for qudits is constructed in a following way:
\begin{equation}
CNOT_d = I \oplus X^1 \oplus X^2 \oplus \ldots \oplus X^{d-2} \oplus X^{d-1},
\end{equation}
where the symbol $\oplus $ stands for matrix direct sum and $X$ means the negation operation for qudit with freedom level $d$.

\begin{figure}
\begin{tabular}{c}
\includegraphics[width=10.75cm]{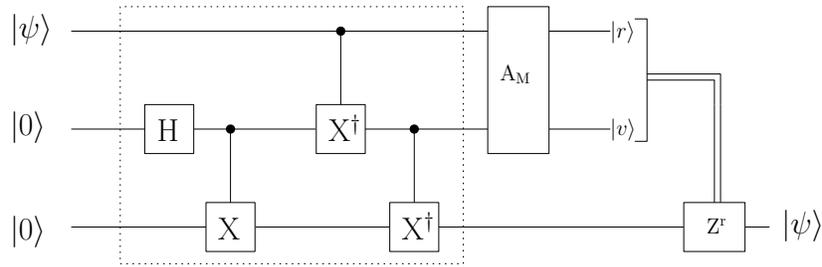} 
\end{tabular}
\centering
\caption{A quantum circuit realizing the quantum teleportation protocol. The correction is performed only by the $Z$ gate. The symbols $X$ and $X^{\dagger}$ stand for CNOT-type gates. The gate $A_M$ takes part in a process of measurement on the first and the second qudit, but to perform the correction operation only the value of first qudit's amplitude is necessary}
\label{lbl:fig:circuit:teleporation:MS:JW:CN:2016}
\end{figure}

\begin{figure}
	\begin{tabular}{c}
		\includegraphics[width=10.75cm]{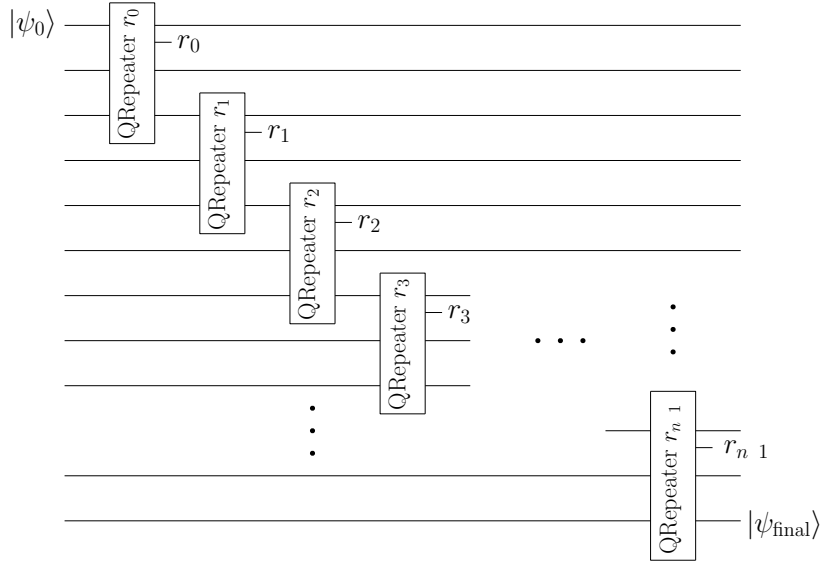}
	\end{tabular}
	\centering
	\caption{The general scheme of quantum circuit which realizes the one-way communication protocol. There are $3 \cdot n $ qudits needed to send the information between the initial and the final node. It is assumed that the error correction is performed locally on each repeater. In this one-way protocol the communication is organized from qudit marked as $\mket{\psi_0}$ to qudit $\mket{\psi_{final}}$}
	\label{lbl:fig:one-way-quantum-repeater:MS:JW:CN:2016}
\end{figure}

The gates, mentioned above, allow to present the teleportation protocol of an unknown quantum state:
\begin{equation}
\mket{\psi}=\alpha_0\mket{0} + \alpha_2\mket{1} + \alpha_2\mket{2} + \ldots + \alpha_{d-1}\mket{d-1} \;\;\; \mathrm{and} \;\;\; \sum_{i=0}^{d-1} {|\alpha_i|}^2 = 1 . \label{lbl:eq:unknown_teleported_state:one:dit}
\end{equation}
The 3-qudit state $\mket{\psi A B}$, where $A$ belongs to Alice and $B$ to Bob, after performing the Hadamard, CNOT and $CNOT^{\dagger}$ operations on it, due to Fig.~\ref{lbl:fig:circuit:teleporation:MS:JW:CN:2016}, may be described as:
\begin{equation}
\begin{aligned}
\mket{00}\left(\omega^{F(0,0)}\alpha_0\mket{0} + \omega^{F(0,1)}\alpha_1\mket{1} + \omega^{F(0,2)}\alpha_2\mket{2} + \ldots + \omega^{F(0,w)}\alpha_w\mket{w} \right) + \\ 
\mket{01}\left(\omega^{F(0,0)}\alpha_0\mket{0} + \omega^{F(0,1)}\alpha_1\mket{1} + \omega^{F(0,2)}\alpha_2\mket{2} + \ldots + \omega^{F(0,w)}\alpha_w\mket{w} \right) + \\
+ \ldots\ldots\ldots\ldots\ldots\ldots\ldots\ldots\ldots\ldots\ldots\ldots\ldots + \\ 
\mket{0w}\left(\omega^{F(0,0)}\alpha_0\mket{0} + \omega^{F(0,1)}\alpha_1\mket{1} + \omega^{F(0,2)}\alpha_2\mket{2} + \ldots + \omega^{F(0,w)}\alpha_w\mket{w} \right) + \\
+\ldots\ldots\ldots\ldots\ldots\ldots\ldots\ldots\ldots\ldots\ldots\ldots\ldots + \\ 
\mket{w0}\left(\omega^{F(w,0)}\alpha_0\mket{0} + \omega^{F(w,1)}\alpha_1\mket{1} + \omega^{F(w,2)}\alpha_2\mket{2} + \ldots + \omega^{F(w,w)}\alpha_w\mket{w}\right) + \\
\mket{w1}\left(\omega^{F(w,0)}\alpha_0\mket{0} + \omega^{F(w,1)}\alpha_1\mket{1} + \omega^{F(w,2)}\alpha_2\mket{2} + \ldots + \omega^{F(w,w)}\alpha_w\mket{w}\right) + \\
\mket{w2}\left(\omega^{F(w,0)}\alpha_0\mket{0} + \omega^{F(w,1)}\alpha_1\mket{1} + \omega^{F(w,2)}\alpha_2\mket{2} + \ldots + \omega^{F(w,w)}\alpha_w\mket{w}\right) + \\
+\ldots\ldots\ldots\ldots\ldots\ldots\ldots\ldots\ldots\ldots\ldots\ldots\ldots + \\ 
\mket{ww}\left(\omega^{F(w,0)}\alpha_0\mket{0} + \omega^{F(w,1)}\alpha_1\mket{1} + \omega^{F(w,2)}\alpha_2\mket{2} + \ldots + \omega^{F(w,w)}\alpha_w\mket{w}\right)
\end{aligned}
\end{equation}
where $w=d-1$. The gates CNOT and $CNOT^{\dagger}$ realize the entanglement. Function $F$ is expressed as:
\begin{equation}
F(a,b) = (d - a \cdot b) \; \mathrm{mod} \; d .
\end{equation}
A correction gate's form depends on the result of measurement performed on the first qudit by Alice (however, both qudits are measured in the standard base what is a consequence of Hadamard gate's use in a quantum cirucit -- see Fig.~\ref{lbl:fig:one-way-quantum-repeater:MS:JW:CN:2016}). If Alice, after the measurement, obtains one of given states
\begin{displaymath}
\mket{00}, \mket{01}, \mket{02}, \ldots, \mket{0w}, 
\end{displaymath}
then the Bob's qudit is characterized by the proper values of probability amplitudes and there is no additional operations needed to correct the obtained state. This also stands with the accordance to the values of function $F$: $F(0,1)$, $F(0,2), \ldots$ - they equal zero.

If Alice, also after the operation of measurement, obtains one of following states
\begin{displaymath}
\mket{10}, \mket{11}, \mket{12}, \ldots, \mket{1w},
\end{displaymath}
then the state of Bob's qudit is
\begin{equation}
\omega^0\alpha_0\mket{0} + \omega^w\alpha_1\mket{1} + \omega^{w-1}\alpha_2\mket{2} + \ldots + \omega^{1}\alpha_w\mket{w}.
\end{equation}
The values of $F$ function in this case are following:
\begin{align}
F(1,0)=(d - 1 \cdot 0) \; \mathrm{mod} \; d & = d \; \mathrm{mod} \; d = 0, \\ \notag
F(1,1)=(d - 1 \cdot 1) \; \mathrm{mod} \; d & = d-1 \; \mathrm{mod} \; d = d-1 = w, \\ \notag
F(1,2)=(d - 1 \cdot 2) \; \mathrm{mod} \; d & = d-2 \; \mathrm{mod} \; d = d-2 = w - 1, \\ \notag
\vdots \\ \notag
F(1,w)=(d - 1 \cdot w) \; \mathrm{mod} \; d & = d-(d-1) \; \mathrm{mod} \; d = 1. \\ \notag
\end{align}
When the $Z$ gate is used only once, then the correct values of some amplitudes may be recovered. In general, the result of measurement determines the number of operations $Z$ which should be performed. It can be expressed as $Z^r$ if $r$ stands for the result of measurement performed on the first qudit. 

The above description presents the teleportation protocol for an unknown qudit, where $Z$ gate is needed to correct the final state and only one measurement decides about the way how the correction is done after the teleportation process.

\section{Protocol for quantum repeater} \label{lbl:sec:protocol:quantum:repeater:MS:JW:CN:2016}

\begin{figure}[!h]
	\begin{tabular}{c}
		\includegraphics[width=10.75cm]{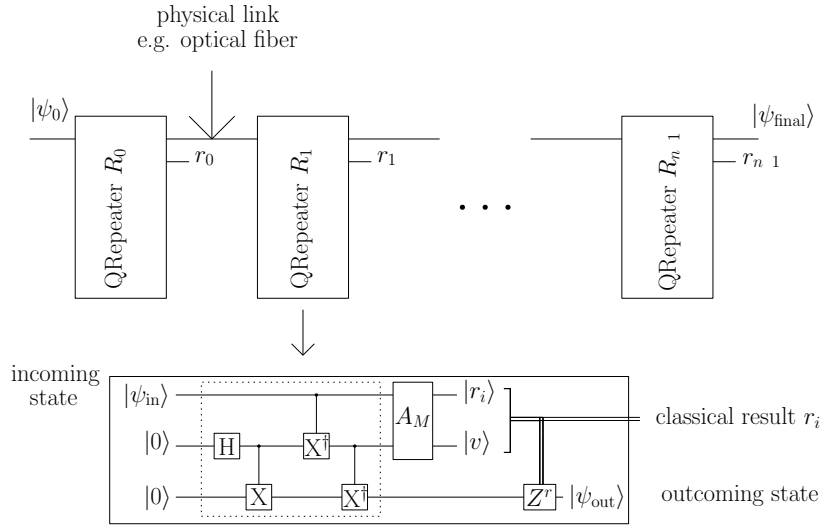}
	\end{tabular}
	\centering
	\caption{The general scheme for the one-way communication protocol based on a teleportation protocol and a quantum repeater. The appearing Pauli errors are corrected with a generalized gate $Z$ (QRepeater -- Quantum Repeater)
	}
\label{lbl:fig:general:scheme:MS:JW:CN:2016}
\end{figure}

The presented protocol of quantum communication is based on the idea of teleportation. It belongs to the third generation of protocols utilizing the concept of repeater. It is so-called one-way protocol and the information is transferred from left to right -- according to Fig.~\ref{lbl:fig:general:scheme:MS:JW:CN:2016}.

The state $\mket{\psi_0}$ is the initial state of the system and it is transferred between repeaters with use of some physical medium. The final state is presented by a vector $\mket{\psi_{final}}$. The changes of system's state, after passing through following repeaters, may be described as:
\begin{equation}
\mket{\psi_0} \overset{r_1}{\rightarrow} \mket{\psi_1} \overset{r_2}{\rightarrow} \mket{\psi_2} \overset{r_2}{\rightarrow} \ldots \overset{r_{n-2}}{\rightarrow} \mket{\psi_{n-1}} \overset{r_{n-1}}{\rightarrow} \mket{\psi_n} \overset{r_n}{\rightarrow} \mket{\psi_{final}}
\end{equation}
All states, from $\mket{\psi_1}$ to $\mket{\psi_{final}}$, need the correction of transferred information because of the protocol's structure. This action may be performed in two ways. The first solution is to correct the state $r_i$ (i.e. the result of measurement performed on Alice's qudit) locally in each repeater by using the gate $ Z^{r_i}$. In this case the local operations guarantee that the state $\mket{\psi_{final}}$ is equal to the initial state in terms of Fidelity measure:

\begin{equation}
F( \mket{\psi_0}, \mket{\psi_{final}} ) = \mbraket{\psi_0}{\psi_{final}} = 1
\end{equation}
where $F$ represent the Fidelity measure and the states $\mket{\psi_0}$, $\mket{\psi_{final}}$ are pure.

The second approach needs to collect all the results of local measurement from every repeater:
\begin{equation}
R = \{ r_1, r_2, r_3, \ldots, r_{n-1}, r_{n}  \}
\end{equation}
The obtained results should be sent to the last node involved in the communication protocol. In the last node the final correction may be applied according to the following formula:
\begin{equation}
\mket{\psi_{Final}} = Z^f \mket{\psi_{final}}  \;\;\; \mathrm{where} \;\;\; f = \left( \sum_{i=1}^{n} R_i \right) \mod d.
\end{equation}
where $d$ is a qudit's freedom level and the state on the last repeater is $\mket{\psi_{final}}$.

\section{Quantum repeaters network as a quantum circuit} \label{lbl:sec:qc:for:qr:network:MS:JW:CN:2016}

Just like in the case of one-dit teleportation protocol, the data transmission with use of QR may be presented as a quantum circuit. The schema of circuit for the teleportation protocol is shown in Fig.~\ref{lbl:fig:circuit:teleporation:MS:JW:CN:2016}. The construction of circuit for quantum transmission needs a system consisting of $3 \cdot n$ qudits and $n$ is the number of used repeaters.

Although the suggested number of qudits needed to carry a correct simulation is $3 \cdot n $, it should be emphasized that the phenomenon of entanglement occurs inside of repeaters. It means that the entanglement is present in a teleportation process and it occurs between three qudits which consist the repeater itself. There is no entanglement between particular repeaters.

This assumption makes easier the implementation of mentioned quantum phenomena with use of classic computers. It is also important to know that the repeaters are connected physically (e.g. optical fiber) with suitable quantum channel.  If the numeric or symbolic simulation is to be carried out on a classic machine, then only the three-qudit repeater needs to be simulated. The features of communication channel, which transmits a quantum state, may be simulated apart from the simulation of QR.

\begin{figure}
\begin{lstlisting}[mathescape]
-- N : numer of nodes
-- i : index of node
-- $\mket{\psi_0}$ : initial state
-- H : change history of the transmitted state

H $\leftarrow$ append(H, $\mket{\psi_0}$, 0)
[$\mket{\psi_t}$, r] $\leftarrow$ QRepeater( $\mket{\psi_0}$ )
H $\leftarrow$ append(H, $\mket{\psi_t}$, r)
$\mket{\psi_t}$ $\leftarrow$ $Z^r \mket{\psi_t}$
while i < N do
begin
	[$\mket{\psi_t}$, r] $\leftarrow$ QRepeater( $\mket{\psi_t}$, r )
	...
	other operations e.g. noise generation
	entanglement level measuring and etc.
	...
	H $\leftarrow$ append(H, $\mket{\psi_t}$, r)
	$\mket{\psi_t}$ $\leftarrow$ $Z^r \mket{\psi_t}$	
	i $\leftarrow$ i + 1
end
$\mket{\psi_{final}}$ $\leftarrow$ $\mket{\psi_t}$
\end{lstlisting} 
	\centering
	\caption{The algorithm for classic simulation of quantum transmission protocol based on $N$ nodes with QRs. The symbol $\leftarrow$ stands for the classic assignment operation}
	\label{lbl:fig:classical-sim-of-quantum-repeater:MS:JW:CN:2016}
\end{figure}
Fig.~\ref{lbl:fig:classical-sim-of-quantum-repeater:MS:JW:CN:2016} presents the pseudo-code illustrating the simulation's algorithm for a given state. The changes are saved in a classic variable $H$ which collects the quantum states obtained after the calculations in every node. The initial state is assigned to variable $\mket{\psi_0}$ and the final state to $\mket{\psi_{final}}$.

\section{Conclusions} \label{lbl:sec:conclusions:MS:JW:CN:2016}

The one-dit teleportation protocol with error correction performed by Pauli-Z gate allows the realization of QR for transmitting the unknown qudit state. Two presented strategies for Z-type error correction provide the choice to the user who can run the protocol with the error correction at the end of the whole process or the correction may be performed locally on every QR. The approach with the correction at the end of transmisssion reduces the use of Z gate application. We use this gate only at the end of the process to complete the communication protocol by correcting the received information. This solution excludes the need of using Pauli-X and Hadamard gates, which are utilized in other known protocols for QR realization. The advantages connected with presented approach are: the influence of possible errors, caused by the imperfect realization of quantum gates, is lower and the construction of QR is easier because only the operation of measurement have to be performed (the error correction is redundant).

Planned further research on presented protocol is to use it as a quantum key distribution protocol. In this case it will be very important to estimate the number of secret bits one can generate.

The another interesting problem for further research is using quantum error correcting codes on transmitted state. The presented protocol concerns the pure states, so it would be also important to analyze the quality of transmitted information taking into account the quality of utilized quantum gates and the influence of environment.

\subsubsection*{Acknowledgments}

We would like to thank for useful discussions with the~\textit{Q-INFO} group at the Institute of Control and Computation Engineering (ISSI) of the University of Zielona G\'ora, Poland. We would like also to thank to anonymous referees for useful comments on the preliminary version of this paper. The numerical results were done using the hardware and software available at the ''GPU $\mu$-Lab'' located at the Institute of Control and Computation Engineering of the University of Zielona G\'ora, Poland.

\end{document}